\def\be{\begin{equation}}
\def\ee{\end{equation}}
\def\ba{\begin{array}}
\def\ea{\end{array}}

\documentclass[11pt]{article}
\usepackage{amsfonts}
\usepackage{amssymb}
\usepackage{dsfont}
\usepackage{amsmath,amssymb,amstext,amsfonts,array,hhline,tabularx,graphics}
\usepackage{amsthm}

\usepackage{latexsym}
\usepackage{arydshln}
\usepackage{epsfig,graphicx, color}
\usepackage{amsmath}
 \usepackage{cite}
\usepackage{amsthm}
\usepackage{braket}
\graphicspath{{photography/}}
\usepackage{subcaption}
\usepackage{graphicx}
\usepackage{pgfplots}
\usepackage{inputenx}
\usepackage{epstopdf}

\theoremstyle{plain}
\begin{document}
\parskip=3pt
\parindent=18pt
\baselineskip=20pt \setcounter{page}{1}

 \title{\large\bf Quantum coherence of mixed states under noisy channels in noninertial frames}
\date{}

\author{Tangrui Liao$^{1}$, Junhao Yang$^{1}$, Tinggui Zhang$^{1}$, Xiaofen Huang$^{^\ast,  1, 2}$ \\[10pt]
\footnotesize
\small 1 School of Mathematics and Statistics, Hainan Normal University, Haikou 571158, China\\
\small 2 Key Laboratory of Data Science and Smart Education, \\
\small  Ministry of Education, Hainan Normal
University, Haikou 571158, China\\
}
\date{}

\maketitle

\centerline{$^\ast$ Correspondence to  huangxf1206@163.com }
\bigskip

\begin{abstract}
We focus our attention on tripartite mixed states as initial states, and apply coherence concurrence to investigate quantum coherence properties in the background of a Schwarzschild black hole under phase damping, phase flip and bit flip channels, respectively. Several analytic complementary relationships based on coherence concurrence for tripartite subsystems are proposed. In the case of the bit flip channel, the behavior of the coherence concurrence is similar to the one of the phase damping channel, the accessible coherence concurrence always degrades as the Hawking acceleration rising, but sudden death never occurs, while the inaccessible coherence increases from zero monotonically. Interestingly, the coherence concurrence is decreasing at first and then increasing as the decay probability rising under phase flip channel. Unlike the case of tripartite pure states, the coherence concurrence of mixed state with X shape is equal to $l_1$ -norm of coherence.

\end{abstract}

$a'$

\section{Introduction}

Quantum coherence is an important feature in quantum physics and is of practical significance in quantum computation and quantum communication \cite{app1, app2, app3}.  Like entanglement and discord-type quantum correlation, quantum coherence is also a valuable resource for
quantum information processing tasks, quantum thermodynamics, life sciences and condensed matter physics.

The formulation of the resource theory of coherence was initiated in Ref. \cite{coh}, in which some intuitive and computable measures of coherence are identified, such as the $l_1$ -norm coherence and the relative entropy coherence. Given an orthogonal basis $\{\ket{i}\}$, a quantum state is called incoherent if it is diagonal under this basis, i.e., $\rho=\sum_i\rho_i \ket{i}\bra{i}$. Otherwise, it is  said to be coherent.
$l_1$ -norm  coherence is one of the most popular coherence measure. For a quantum state $\rho=\sum \rho_{ij}\ket{i}\bra{j}$,  the $l_1$ -norm coherence of $\rho$ can be denoted as the sum of the magnitudes of all off-diagonal entries, that is,
\begin{equation}\label{l1}
C_{l_1}(\rho)=\sum_{i\neq j} |\rho_{ij}|.
\end{equation}
Obviously, $l_1$ -norm coherence measure in Eq. (\ref{l1}) depends on the choice of orthogonal basis.

However, coherence concurrence, as another quantum coherence measure, is independent of choice of orthogonal basis for a mixed state.
For a $d$-dimensional pure state $\ket{\psi}$, we define its coherence concurrence as
\begin{equation}
\mathcal{C}(\ket{\psi})=\sum_{1\leq j<k\leq d}|\langle \psi| \Lambda_{j k} | \psi^{\ast} \rangle|,
\end{equation}
where $\ast$ denotes complex conjugation, and $\Lambda_{j k}=\ket{j}\bra{k}+\ket{k}\bra{j}$ is a symmetric Hermitian operator.
Furthermore, coherence concurrence can be extended to mixed state by convex roof construction, which is given by \cite{gao2017}
\begin{equation}
\mathcal{C}(\rho)=\mathop{min}\limits_{\{p_i, \ket{\psi_i}\}}\sum p_i\mathcal{C}(\ket{\psi_i}),
\end{equation}
where the minimization is taken over all possible ensemble realizations $\rho=\sum_ip_i\ket{\psi_i}\bra{\psi_i}$, while $p_i\geq 0$ and $\sum_ip_i=1$.
It notes that the coherence concurrence equals $l_1$ -norm of coherence for pure states, for qudit mixed states, $l_1$ -norm of coherence is no more than the coherence concurrence \cite{gao2017}.
Exactly, the essential reason is that the $l_1$ -norm of coherence for a pure state is basis dependent.



On the other hand, as a combination of quantum information, quantum field theory and relativity theory,
relativistic quantum information has attracted considerable attention
for a long time \cite{mann, alsing, aspachs, tian2017, tian2018, unrh3}.
It is obvious that these investigations in a relativistic
framework are not only helpful in understanding some of the key questions in quantum
information, but also play an important role in the study of the quantum coherence
and information paradox of black holes \cite{hawking1975, hawking1976}.

In recent years, much effort has been
given to understand how the Unruh effect or Hawking effect influences the degree of  quantum coherence of tripartite states in non-inertial frames \cite{wu2019, wu2021, wz2021, har2021, kim, unrh1, unrh2}.
In Ref. \cite{wu2019}, authors studied the quantum coherence of GHZ-like states of
multimode Dirac fields in the background of a Schwarzschild black hole. Wu \textit{et al} \cite{wu2021} investigated the quantum
coherence for tripartite GHZ and W states under the amplitude-damping environment in non-inertial frames. In Ref. \cite{wz2021}, the quantum coherence and its distribution of N-partite GHZ and W states of bosonic fields in non-inertial frames are discussed. In Ref. \cite{har2021}, the authors investigated the quantum coherence of the GHZ and W states when some of the parties are moving with constant acceleration. They found that the quantum coherence of these tripartite systems does not vanish in the infinite acceleration limit and derived analytic expressions for the quantum coherence of N-partite GHZ and W states in non-inertial frames.

In this work, we discuss the properties of multipartite coherence for Dirac fields in
the background of a Schwarzschild black hole. To illustrate the problem conveniently,
we first assume Alice, Bob and Charlie share a tripartite mixed state in flat Minkowski spacetime,
and then, let Alice stay stationary at an
asymptotically flat region while Bob and Charlie hover near the event horizon of the
black hole with uniform acceleration. We focus our attention on the effect of Hawking
radiation and interaction with noise to the accessible and inaccessible coherence concurrence.

\section{Coherence concurrence under the effect of Hawking radiation}

This article primarily employs coherence concurrence as a measure of quantum coherence of mixed states under effects of noise channels and accelerating in noninertial frames.
Generally, for any mixed state, it is a challenge to compute coherence concurrence.
However, for any $n$ dimensional quantum state $\rho$ with density matrix in X shape, i.e.,
\begin{equation}
	\rho=\left(\begin{array}{ccccccc}
		\rho_{11} & 0 & 0 & \cdots & 0 & 0 & \rho_{1, n} \\
		0 & \rho_{22} & 0 & \cdots & 0 & \rho_{2, n-1} & 0 \\
		0 & 0 & \rho_{33} & \cdots & \rho_{3, n-2} & 0 & 0 \\
		\cdots & \cdots & \cdots & \cdots & \cdots & \cdots & \cdots \\
		0 & 0 & \rho_{n-2,3} & \cdots & \rho_{n-2, n-2} & 0 & 0 \\
		0 & \rho_{n-1,2} & 0 & \cdots & 0 & \rho_{n-1, n-1} & 0 \\
		\rho_{n, 1} & 0 & 0 & \cdots & 0 & 0 & \rho_{n n}
	\end{array}\right) ,
\end{equation}
the measure of coherence concurrence utilized for the computation of the X state is delineated as follows \cite{zhao2020},
	\begin{equation}\label{concur}
		\mathcal{C}(\rho)=2 \sum_{i=1}^{[n / 2]}|\rho_{i, n+1-i}|,
	\end{equation}
where $[n / 2]$ denotes the integer part of $n$.

We consider the scenario for measuring  coherence concurrence involves three parties, Alice(A), Bob(B) and Charlie(C). They share a three-qubit quantum system with expression provided by
\begin{equation}\label{MGHZ}
 	\rho=\alpha\ket{GHZ}\bra{GHZ}+(1-\alpha)\ket{000}\bra{000},
 \end{equation}
where the state parameter $\alpha$ runs from 0 to 1.
Now we assume that Alice stays at an asymptotically flat region, while Bob and Charlie hover with an uniform acceleration near the event horizon of a
 Schwarzschild black hole. Due to the Hawking radiation of black hole, the Dirac fields observed by Bob and Charlie will change.
 In the single-mode approximation, the vacuum and one-particle states in flat Minkowski spacetime are transformed into \cite{unruh}
 \begin{equation}\label{trans1}
|0\rangle_M=\cos r|0\rangle_{\textit{I}}|0\rangle_{\textit{II}}+\sin r|1\rangle_{\textit{I}}|1\rangle_{\textit{II}},
\end{equation}
and
\begin{equation}\label{trans2}
|1\rangle_M=|1\rangle_{\textit{I}}|0\rangle_{\textit{II}},
\end{equation}
where the acceleration parameter $r$ is defined by $\cos r=(e^{-2\pi \omega c/a}+1)^{-1/2}$ while $0 \leq r\leq \pi/4$, $a$ is the acceleration of the accelerated observer, while $\omega$ is frequency of the Dirac particle, $c$ is the speed of light in vacuum. $\{ |n\rangle_{\textit{I}(\textit{II})}\} (n=0, 1)$  indicates Rindler modes in region $\textit{I}(\textit{II})$.

In the following, we study the effect of Hawking radiation on the quantum  coherence for different subsystems with measure of coherence concurrence.
Since the interior region is causally disconnected from the exterior region of the Schwarzschild black hole, we call the modes inside the event horizon, i. e., the $B_{II}$ and $C_{II}$, the inaccessible modes and the modes outside the event horizon ( $B_{I}$ and $C_{I}$ ) the accessible modes. Since Bob and Charlie can not access to the Rindler region $II$, we should take trace over modes $ B_{\textit{II}}$ and $ C_{\textit{II}}$. Employing transformation (\ref{trans1}) and (\ref{trans2}) and tracing out the inaccessible modes, we obtain the partial trace quantum state $\rho_{AB_{\textit{I}}C_{\textit{I}}}$ as follows,
\begin{equation}\label{ab1c1}
\rho_{AB_IC_I}=\left(
  \begin{array}{cccccccc}
    a_1 & 0 & 0 & 0 & 0 & 0 & 0 & b_1 \\
    0 & a_2 & 0 & 0 & 0 & 0 & 0 & 0 \\
    0 & 0 & a_3 & 0 & 0 & 0 & 0 & 0 \\
    0 & 0 & 0 & a_4 & 0 & 0 & 0 & 0 \\
    0 & 0 & 0 & 0 & 0 & 0 & 0 & 0 \\
    0 & 0 & 0 & 0 & 0 & 0 & 0 & 0 \\
    0 & 0 & 0 & 0 & 0 & 0 & 0 & 0 \\
    b_1 & 0 & 0 & 0 & 0 & 0 & 0 & \frac{\alpha}{2} \\
  \end{array}
\right),
\end{equation}
where the matrix elements are given by
\begin{equation}
\begin{aligned}
a_1&=\frac{2-\alpha}{2}\cos^2 r_b \cos^2 r_c,\\
a_2&=\frac{2-\alpha}{2}\cos^2 r_b \sin^2 r_c,\\
a_3&=\frac{2-\alpha}{2}\sin^2 r_b \cos^2 r_c,\\
a_4&=\frac{2-\alpha}{2}\sin^2 r_b \sin^2 r_c,\\
b_1&=\frac{\alpha}{2}\cos r_b \cos r_c.\\
\end{aligned}
\end{equation}

In a similarly way, we can calculate other physically inaccessible quantum tripartite subsystems (see details in Appendix A).
According to the analytical formula of coherence concurrence as Eq. (\ref{concur}), we obtain the coherence concurrence for both physically accessible and physically inaccessible subsystems as
\begin{equation}\label{con}
\begin{aligned}
\mathcal{C}(\rho_{AB_IC_I})&=\alpha \cos r_b \cos r_c,\\
\mathcal{C}(\rho_{AB_{II}C_I})&=\alpha\sin r_b \cos r_c,\\
\mathcal{C}(\rho_{AB_{I}C_{II}})&=\alpha\cos r_b \sin r_c,\\
\mathcal{C}(\rho_{AB_{II}C_{II}})&=\alpha\sin r_b \sin r_c,\\
\mathcal{C}(\rho_{AB_{I}B_{II}})&=\alpha \cos^2 r_b,\\
\mathcal{C}(\rho_{AC_{I}C_{II}})&=\alpha \cos^2 r_c.
\end{aligned}
\end{equation}

These analytic expressions of coherence concurrence as Eq. (\ref{con}), clearly,  depend not only on the state parameter $\alpha$ but also on the  acceleration parameters $ r_b$ and $ r_c$.
Suppose Bob stays at an asymptotically flat region, i. e., $r_b=0$, the coherence concurrence of $\rho_{AB_{II}C_I}$ and $\rho_{AB_{II}C_{II}}$ reduce to zero, while $\mathcal{C}(\rho_{AB_IC_{I}})=\mathcal{C}(\rho_{ABC_{I}})=\alpha \cos r_c$ and $\mathcal{C}(\rho_{AB_IC_{II}})=\mathcal{C}(\rho_{ABC_{II}})=\alpha \sin r_c$. As the acceleration rising, we have seen that the physically accessible coherence concurrence monotonically degrades; at the same time, the physically inaccessible coherence concurrence monotonically increases from zero.
This phenomenon shows that there may be complementarity relations which have permeated theoretical descriptions and the modeling of quantum physical systems, and these relations can help determining some variable of interest concerning a second variable complementary to the first one.

Here, the first complementarity relation which indicates this mutual restrictive relationship between $\mathcal{C}(\rho_{AB_IC_{I}})$and $\mathcal{C}(\rho_{AB_IC_{II}})$, is expressed by
\begin{equation}
\mathcal{C}^2(\rho_{AB_IC_{I}})+\mathcal{C}^2(\rho_{AB_IC_{II}})=\alpha^2.
\end{equation}
Also, for physically inaccessible coherence concurrence $\mathcal{C}(\rho_{AB_IC_{II}})$ and $\mathcal{C}(\rho_{AC_{I}C_{II}})$, they share a complementarity relation as
\begin{equation}
\mathcal{C}^2(\rho_{AB_IC_{II}})+\alpha \mathcal{C}(\rho_{AC_{I}C_{II}})=\alpha^2.
\end{equation}
Additionally, we derive another useful complementarity relation from the coherence concurrence of tripartite system after acceleration from Eq. (\ref{con}),
\begin{equation}
\mathcal{C}^2(\rho_{AB_{I}C_{I}})+\mathcal{C}^2(\rho_{AB_{I}C_{II}})
+\mathcal{C}^2(\rho_{AB_{II}C_{I}})+\mathcal{C}^2(\rho_{AB_{II}C_{II}})=\alpha^2.
\end{equation}
This reflects the total coherence concurrence of physically accessible $\mathcal{ C}(\rho_{AB_{I}C_{I}})$ and physically inaccessible $\mathcal{C}(\rho_{AB_{\mathrm{I}}C_{\mathrm{II}}})$, $\mathcal{C}(\rho_{AB_{\mathrm{II}}C_{\mathrm{I}}})$ and $\mathcal{C}(\rho_{AB_{\mathrm{II}}C_{\mathrm{II}}})$ is a constant.

\section{Coherence concurrence affected by noisy environment}
Additionally, it is posited that each individual subsystem experiences perturbations from a distinct noisy environment, characterized by specific interactions. The action of a noisy environment is described as
\begin{equation}
	\rho \rightarrow \rho^{evo}=\sum_k E_k \rho E_k^{\dagger},
\end{equation}
where $\rho(\rho^{evo})$ is the density matrix of a initial (final) state, $E_k$ ($E_k^{\dagger}$) is the single-qubit Kraus (complex conjugate) operator of the noisy channel. Except the Unruh effect, let's move to discuss the interconnections among the formulas under various noisy environments where both Bob and Charlie involve in, here we focus on phase damping channel, phase flip channel  and bit flip channel \cite{hu1, hu2}as examples.

\textbf{In case of phase damping channel}: First of all,  we will consider the case of the phase damping channel. The single qubit Kraus operators for the phase
damping channel are given by
\begin{equation}
\begin{aligned}
&E_0=\left(\begin{array}{cc}
		1 & 0 \\
		0 & \sqrt{1-P}
	\end{array}\right),
&E_1=\left(\begin{array}{cc}
		0 & 0 \\
		0 & \sqrt{P}
	\end{array}\right),\\
\end{aligned}
\end{equation}
where $P \in [0, 1]$ is a decay probability and in our study we assume that it depends only on time \cite{decay}.

Under the phase damping noise, the tripartite reduced state $\rho_{AB_IC_I}$  evolves to a state as follows
\begin{equation}
\rho^{evo}_{AB_IC_I}=\left(
  \begin{array}{cccccccc}
    a_1 & 0 & 0 & 0 & 0 & 0 & 0 & d_1 \\
    0 & a_2 & 0 & 0 & 0 & 0 & 0 & 0 \\
    0 & 0 & a_3 & 0 & 0 & 0 & 0 & 0 \\
    0 & 0 & 0 & a_4 & 0 & 0 & 0 & 0 \\
    0 & 0 & 0 & 0 & 0 & 0 & 0 & 0 \\
    0 & 0 & 0 & 0 & 0 & 0 & 0 & 0 \\
    0 & 0 & 0 & 0 & 0 & 0 & 0 & 0 \\
    d_1 & 0 & 0 & 0 & 0 & 0 & 0 & \frac{\alpha}{2} \\
  \end{array}
\right),
\end{equation}
where the matrix element
\begin{equation}
d_1=\frac{\alpha}{2}\sqrt{1-P_b}\sqrt{1-P_c}\cos r_b \cos r_c.
\end{equation}

Similarly, we can compute the evolved quantum states of the other tripartite reduced states (see details in Appendix B).  Employing Eq. (\ref{concur}), the coherence concurrences of those states under phase damping noise are given by
\begin{equation}\label{e1}
 \begin{aligned}
\mathcal{C}(\rho^{evo}_{AB_IC_I})&=\alpha\sqrt{1-P_b}\sqrt{1-P_c}\cos r_b \cos r_c,\\
\mathcal{C}(\rho^{evo}_{AB_{II}C_{II}})&=\alpha\sqrt{1-P_b}\sqrt{1-P_c}\sin r_b \sin r_c,\\
\end{aligned}
\end{equation}

\begin{equation}\label{e2}
 \begin{aligned}
\mathcal{C}(\rho^{evo}_{AB_{II}C_I})&=\alpha\sqrt{1-P_b}\sqrt{1-P_c}\sin r_b \cos r_c,\\
\mathcal{C}(\rho^{evo}_{AB_{I}C_{II}})&=\alpha\sqrt{1-P_b}\sqrt{1-P_c}\cos r_b \sin r_c,\\
\end{aligned}
\end{equation}

\begin{equation}\label{e3}
 \begin{aligned}
\mathcal{C}(\rho^{evo}_{AB_IB_{II}})&=\alpha\sqrt{1-P_b}\sqrt{1-P_c}\cos^2 r_b ,\\
\mathcal{C}(\rho^{evo}_{AC_IC_{II}})&=\alpha\sqrt{1-P_b}\sqrt{1-P_c} \cos^2 r_c.\\
\end{aligned}
\end{equation}
From Eq. (\ref{e1}), it implies that there may be a physical symmetry between subsystems B and C under Hawking acceleration and noisy interference.

In the case Bob and Charlie hover with the same uniform acceleration near the event horizon of a
 Schwarzschild black hole and under the same decay probability, i.e., $r_b=r_c=r$ and $P_b=P_c=P$, thus, Eqs. (\ref{e1})-(\ref{e3}) reduce to
 \begin{equation}\label{d}
 \begin{aligned}
 \mathcal{C}(\rho^{evo}_{AB_IC_I})&=\mathcal{C}(\rho^{evo}_{AB_IB_{II}})=\mathcal{C}(\rho^{evo}_{AC_IC_{II}})
 =\alpha(1-P)\cos^2 r,\\
 \mathcal{C}(\rho^{evo}_{AB_{II}C_I})&=\mathcal{C}(\rho^{evo}_{AB_{I}C_{II}})=\frac{\alpha}{2} (1-P) \sin 2r,\\
 \mathcal{C}(\rho_{AB_{II}C_{II}})&=\alpha(1-P) \sin^2 r.\\
 \end{aligned}
 \end{equation}

Eq. (\ref{d}) reflects that the coherence concurrence under noisy evolution depends on both parameters $r$ and $P$ for a given system.  These analytic expressions indicate three basis facts:  the first fact is the coherence concurrence in physical accessible modes $AB_IC_I$ decreases with the Hawking acceleration increasing, but it could not vanish even for the maximal acceleration $r=\frac{\pi}{4}$. On the contrary, the coherence concurrences in physical inaccessible modes $AB_IC_{II}$, $AB_{II}C_I$ and $AB_{II}C_{II}$ which involve three parties A, B and C, increase with the Hawking acceleration increasing. The second one is that physical inaccessible modes $AB_IB_{II}$ and $AC_IC_{II}$ involving only two parties decrease as the acceleration rising. The last fact is all of  the coherence concurrences are monotone decreasing as decay probability $P$ increasing, and happen to zero when $P=1$.

To gain a deeper understanding,
we take coherence concurrence as a function of Hawking acceleration $r$, decay probability $P$ and state parameter $\alpha$. We plot coherence concurrences of tripartite reduced states with various values for parameters.
In Fig.1, we show coherence concurrence $\mathcal{C}(\rho_{AB_{I}C_{I}}^{evo})$, $\mathcal{C}(\rho_{AB_{II}C_{II}}^{evo})$ as the function of Hawking acceleration $r$ for different decay probability $P$ and state parameter $\alpha$  when the system is under the influence of a phase damping channel.
\begin{figure}
	\centering
	\begin{subfigure}[h]{0.32\textwidth}
		\includegraphics[width=\textwidth]{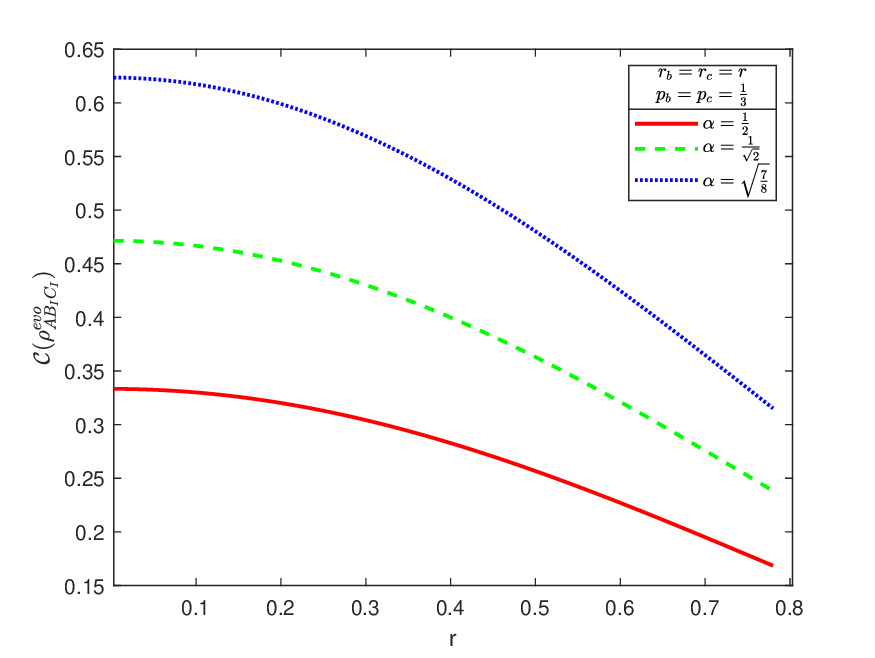}
		\caption{}
		\label{fig:subfig1}
	\end{subfigure}
  \begin{subfigure}[h]{0.32\textwidth}
		\includegraphics[width=\textwidth]{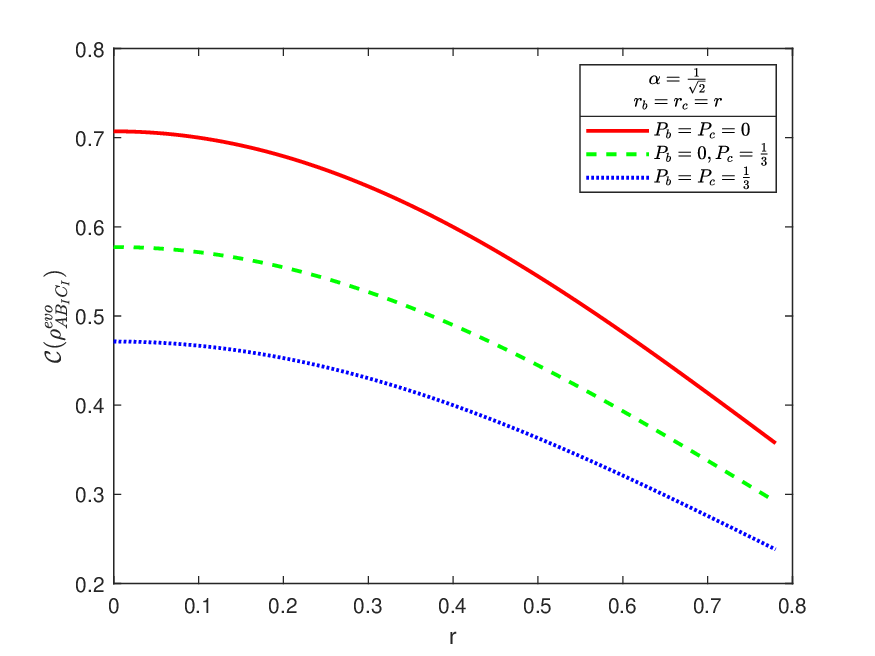}
		\caption{  }
		\label{fig:subfig1}
	\end{subfigure}
 \begin{subfigure}[h]{0.32\textwidth}
		\includegraphics[width=\textwidth]{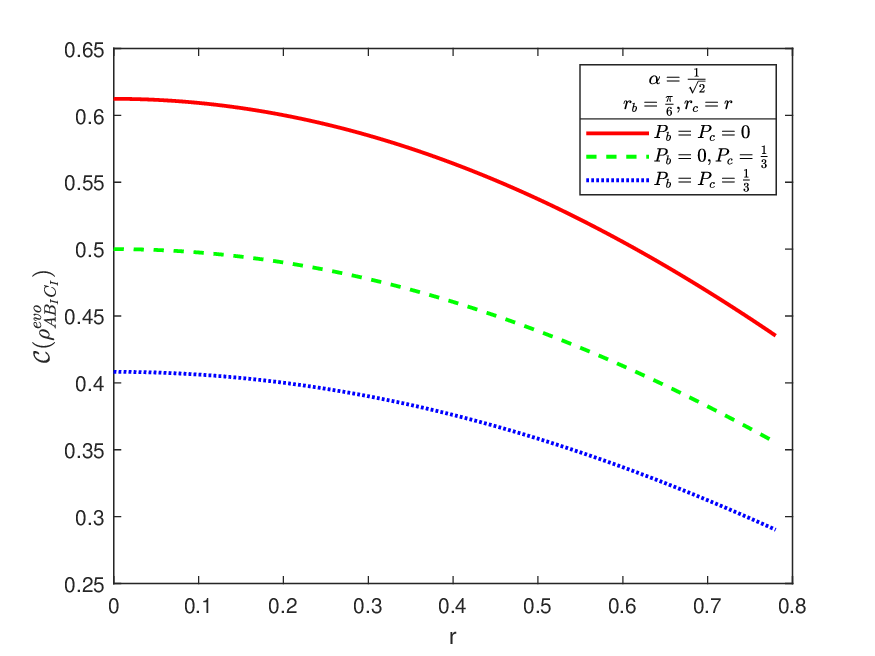}
		\caption{  }
		\label{fig:subfig1}
	\end{subfigure}

\begin{subfigure}[h]{0.32\textwidth}
		\includegraphics[width=\textwidth]{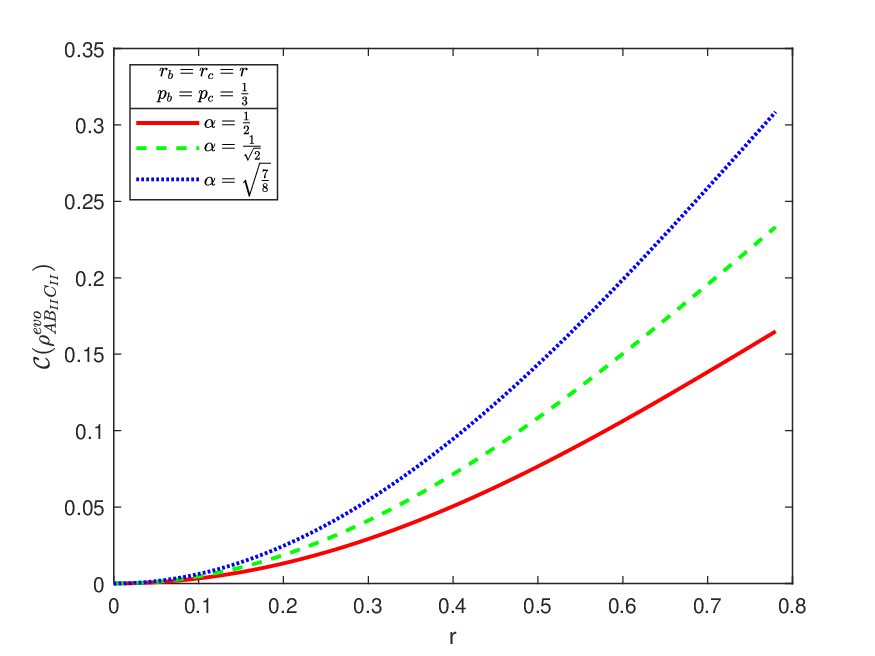}
		\caption{}
		\label{fig:subfig1}
	\end{subfigure}
  \begin{subfigure}[h]{0.32\textwidth}
		\includegraphics[width=\textwidth]{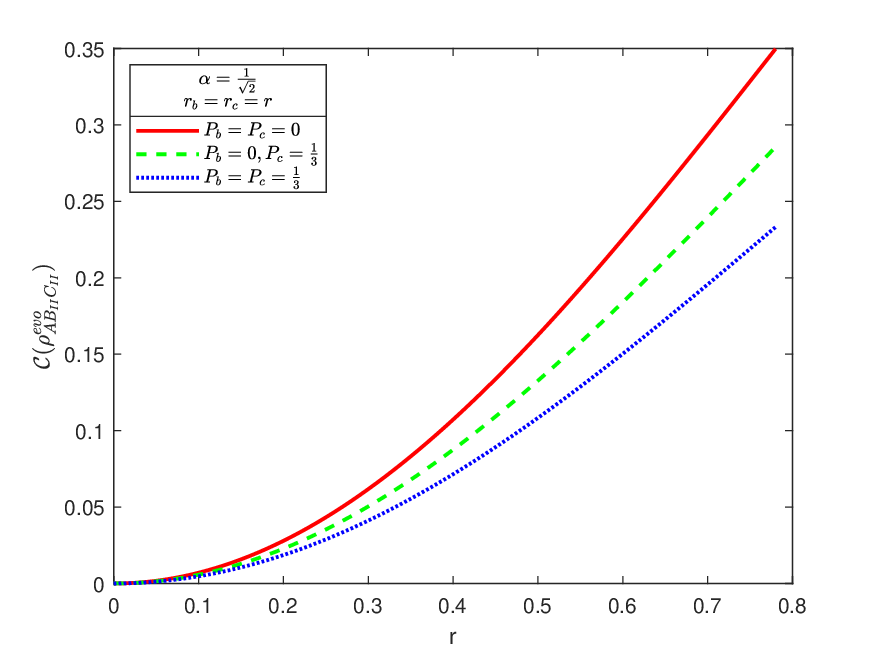}
		\caption{  }
		\label{fig:subfig1}
	\end{subfigure}
 \begin{subfigure}[h]{0.32\textwidth}
		\includegraphics[width=\textwidth]{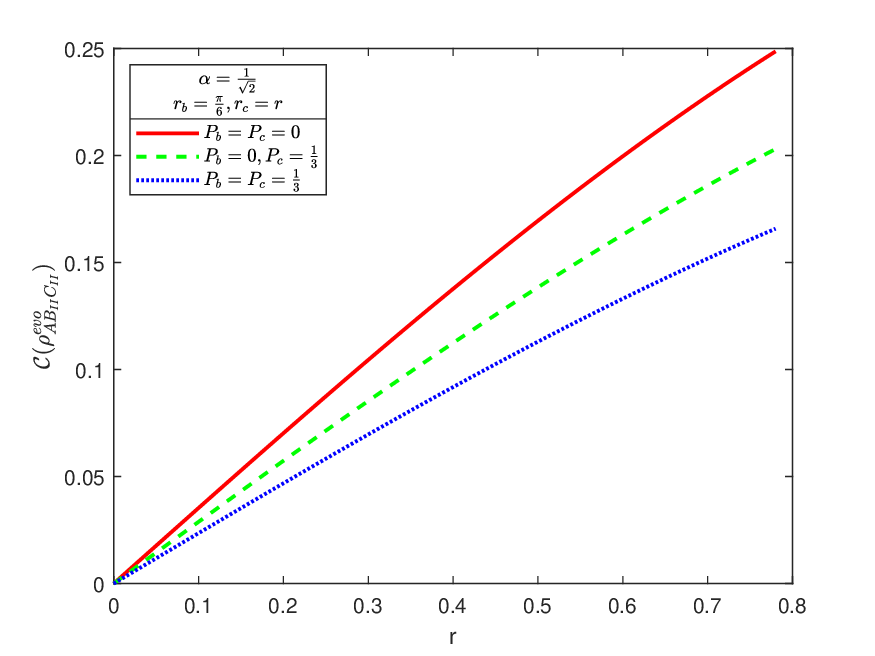}
		\caption{  }
		\label{fig:subfig1}
	\end{subfigure}
	\label{Fig.2}\caption{Plot coherence concurrence $\mathcal{C}(\rho_{AB_{I}C_{I}}^{evo})$  and $\mathcal{C}(\rho_{AB_{II}C_{II}}^{evo})$ as functions of Hawking radiation $r$ under phase damping noise.}
\end{figure}

In Fig.2, we consider the influence of both Hawking radiation and noisy interference simultaneously, therefore, we plot the 3D-figure of coherence concurrences $\mathcal{C}(\rho_{AB_{I}C_{I}}^{evo})$, $\mathcal{C}(\rho_{AB_{II}C_{II}}^{evo})$ and $\mathcal{C}(\rho_{AB_{II}C_{I}}^{evo})$ as the functions of Hawking acceleration $r$ and decay probability $P$ for a fixed system.
\begin{figure}
	\centering
\begin{subfigure}[h]{0.32\textwidth}
		\includegraphics[width=\textwidth]{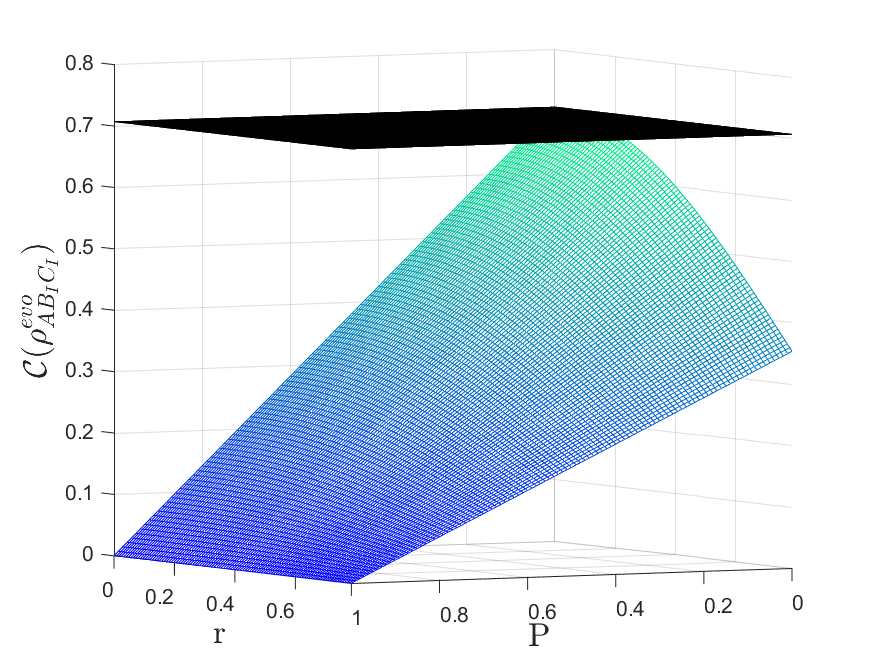}
		\caption{ }
		\label{fig:subfig1}
	\end{subfigure}
	\begin{subfigure}[h]{0.32\textwidth}
		\includegraphics[width=\textwidth]{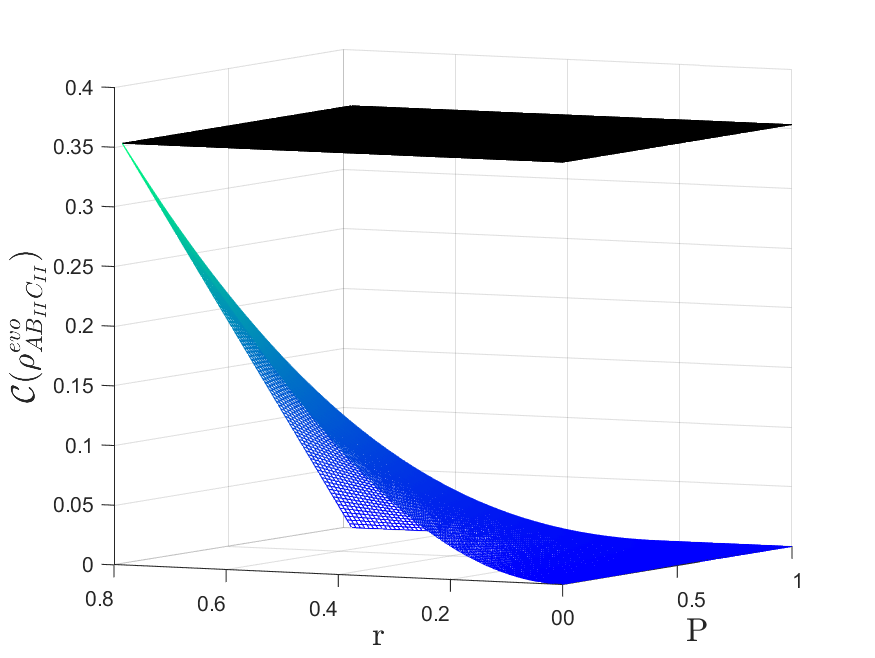}
		\caption{ }
		\label{fig:subfig1}
	\end{subfigure}
  \begin{subfigure}[h]{0.32\textwidth}
		\includegraphics[width=\textwidth]{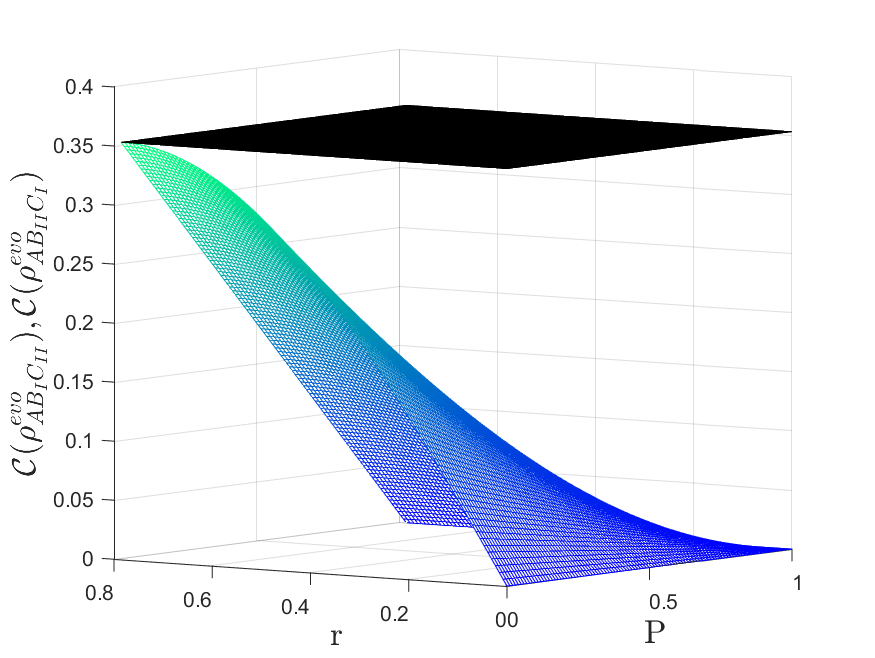}
		\caption{ }
		\label{fig:subfig1}
	\end{subfigure}

	\label{Fig.2}\caption{We plot quantum coherence concurrence $\mathcal{C}(\rho_{AB_{I}C_{I}}^{evo})$, $\mathcal{C}(\rho_{AB_{II}C_{II}}^{evo})$ and $\mathcal{C}(\rho_{AB_{II}C_{I}}^{evo})$ as functions of both decay probability $P$ and Hawking radiation $r$ under phase damping noise  when $r_b=r_c=r$, $P_b=P_c=P$ and $\alpha=\frac{1}{\sqrt{2}}$.}
\end{figure}

These charts illustrate an interesting phenomenon.
With the increasing in Hawking acceleration $r$, the physically accessible quantum coherence concurrence monotonically decrease from a constant value less than 1, however, all the physically inaccessible quantum coherence concurrences are increasing from zero. This interesting phenomenon indicates that the acceleration introduced by Hawking radiation destroys the physically accessible coherence concurrence between Alice, Bob and Charlie. This also implies Hawking radiation is equivalent to a kind of heat effect which reduces the physically accessible coherence and of course may breed the physically inaccessible coherence. Furthermore, from Fig. 1 and Fig. 2, we can see that influence of noisy environment on the  coherence concurrence  is more stronger than the Unruh effect.

\textbf{In case of phase flip channel}:
When we consider the case of the  phase flip channel, the single qubit Kraus operators are given by,
\begin{equation}
\begin{aligned}
&E_0=\left(\begin{array}{cc}
		\sqrt{1-P} & 0 \\
		0 & \sqrt{1-P}
	\end{array}\right),
&E_1=\left(\begin{array}{cc}
		\sqrt{P} & 0 \\
		0 & -\sqrt{P}
	\end{array}\right).\\
\end{aligned}
\end{equation}
Under the joint effects of the acceleration and the phase flip  channel, the initial density matrix
(\ref{ab1c1}) evolves to another form as follows, which reads
\begin{equation}
\rho^{evo}_{AB_IC_I}=\left(
  \begin{array}{cccccccc}
    a_1 & 0 & 0 & 0 & 0 & 0 & 0 & e_1 \\
    0 & a_2 & 0 & 0 & 0 & 0 & 0 & 0 \\
    0 & 0 & a_3 & 0 & 0 & 0 & 0 & 0 \\
    0 & 0 & 0 & a_4 & 0 & 0 & 0 & 0 \\
    0 & 0 & 0 & 0 & 0 & 0 & 0 & 0 \\
    0 & 0 & 0 & 0 & 0 & 0 & 0 & 0 \\
    0 & 0 & 0 & 0 & 0 & 0 & 0 & 0 \\
   e_1 & 0 & 0 & 0 & 0 & 0 & 0 & \frac{\alpha}{2} \\
  \end{array}
\right),
\end{equation}
where element
\begin{equation}
e_1=\frac{\alpha(2 P_b-1 ) (2 P_c-1 )}{2} \cos r_b \cos r_c.
\end{equation}
Therefore, we have
\begin{equation}
\mathcal{C}(\rho^{evo}_{AB_IC_I})=|\alpha(2 P_b-1 ) (2 P_c-1 )| \cos r_b \cos r_c.
\end{equation}

Similarly, we calculate the coherence concurrences for physically inaccessible after influence by the phase flip channel, that is,
\begin{equation}
\begin{aligned}
\mathcal{C}(\rho^{evo}_{AB_IC_{II}})&=|\alpha(2 p_b-1 ) (2 p_c-1 )| \cos r_b \sin r_c,\\
\mathcal{C}(\rho^{evo}_{AB_{II}C_{I}})&=|\alpha(2 p_b-1 ) (2 p_c-1 )| \sin r_b \cos r_c,\\
\mathcal{C}(\rho^{evo}_{AB_{II}C_{II}})&=|\alpha(2 p_b-1 ) (2 p_c-1 )| \sin r_b \sin r_c.\\
\end{aligned}
\end{equation}

In Fig. 3 and Fig. 4, we show the coherence concurrence as functions of Hawking acceleration $r$ and state parameter $\alpha$, when the system is under the influence of a phase flip channel. These figures in Fig. 3 indicate that the coherence concurrence in physically inaccessible mode is decreasing as acceleration rising, but, the coherence concurrence sudden death never occurs. That is similar as the case of the phase damping channel.
\begin{figure}
	\centering
	\begin{subfigure}[h]{0.32\textwidth}
		\includegraphics[width=\textwidth]{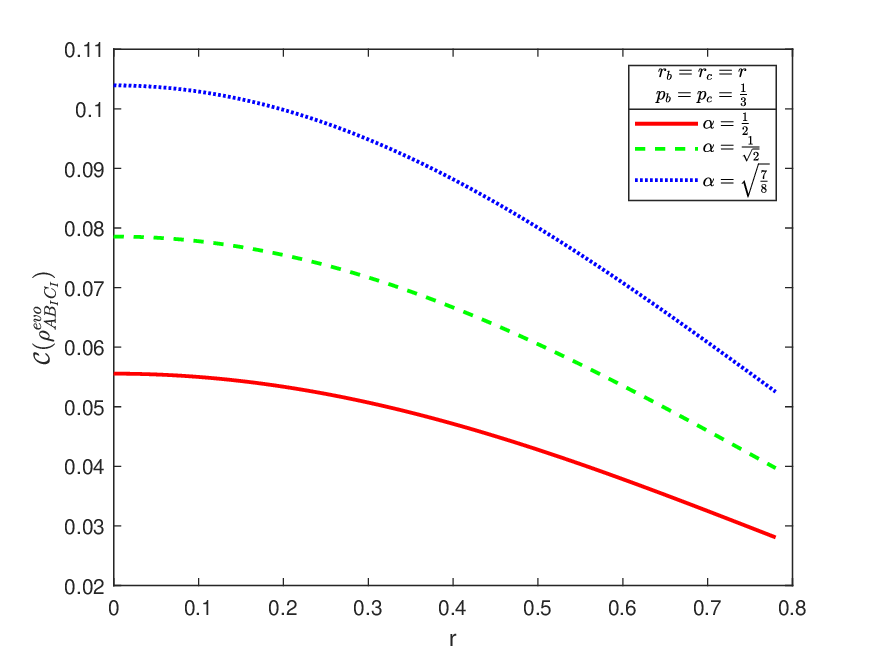}
		\caption{ }
		\label{fig:subfig1}
	\end{subfigure}
  \begin{subfigure}[h]{0.32\textwidth}
		\includegraphics[width=\textwidth]{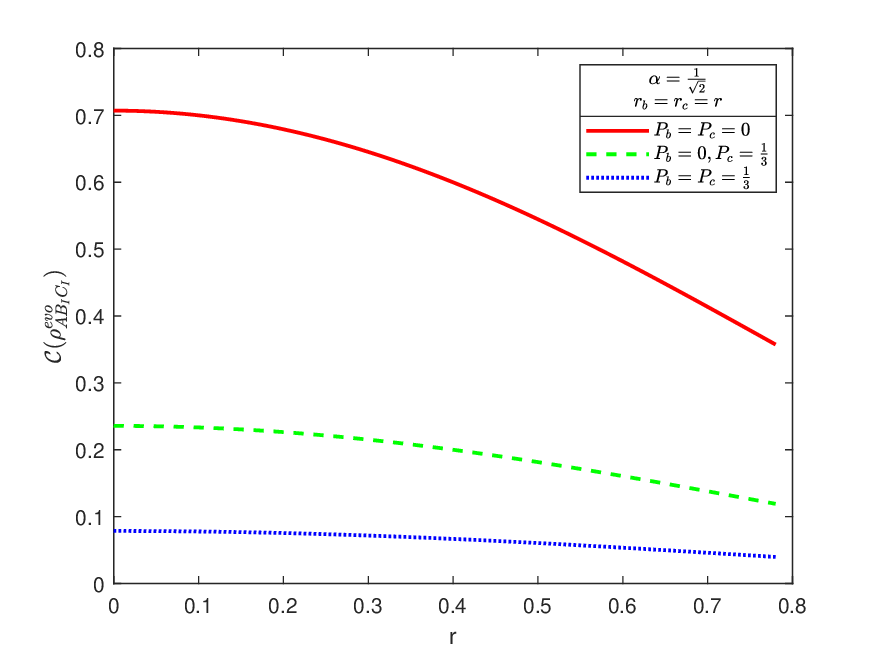}
		\caption{ }
		\label{fig:subfig1}
	\end{subfigure}
 \begin{subfigure}[h]{0.32\textwidth}
		\includegraphics[width=\textwidth]{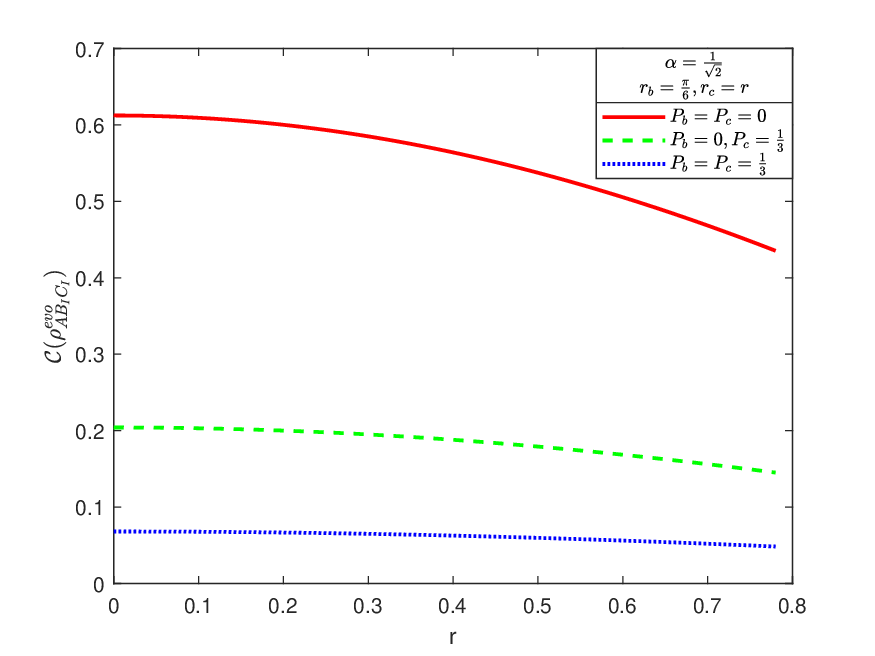}
		\caption{  }
		\label{fig:subfig1}
	\end{subfigure}
	\begin{subfigure}[h]{0.32\textwidth}
		\includegraphics[width=\textwidth]{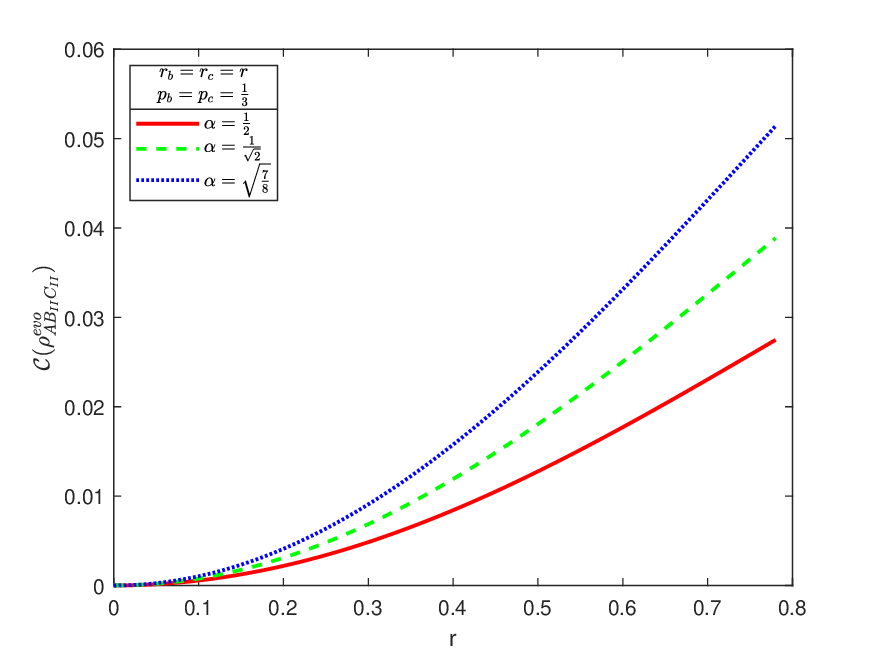}
		\caption{ }
		\label{fig:subfig1}
	\end{subfigure}
  \begin{subfigure}[h]{0.32\textwidth}
		\includegraphics[width=\textwidth]{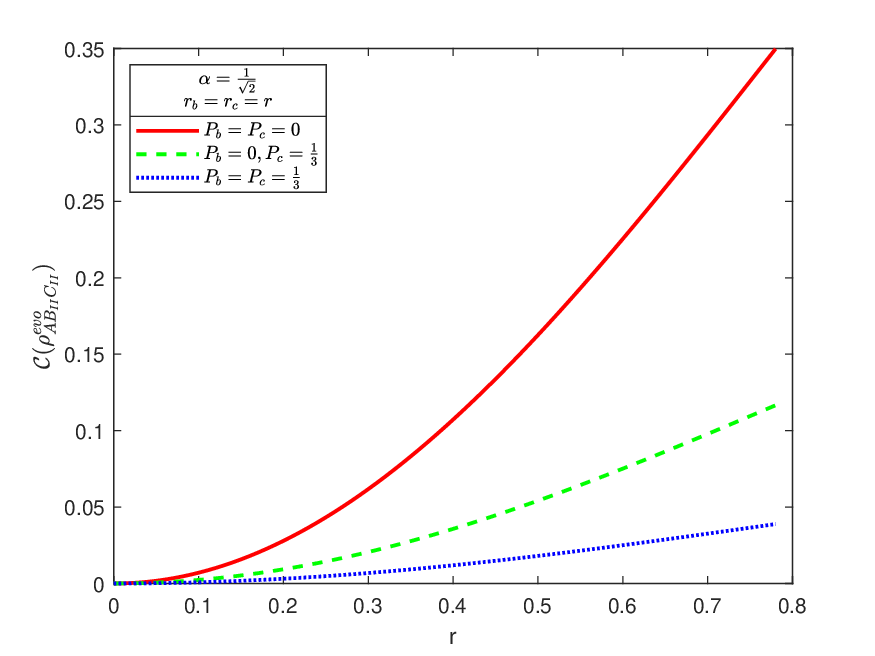}
		\caption{ }
		\label{fig:subfig1}
	\end{subfigure}
 \begin{subfigure}[h]{0.32\textwidth}
		\includegraphics[width=\textwidth]{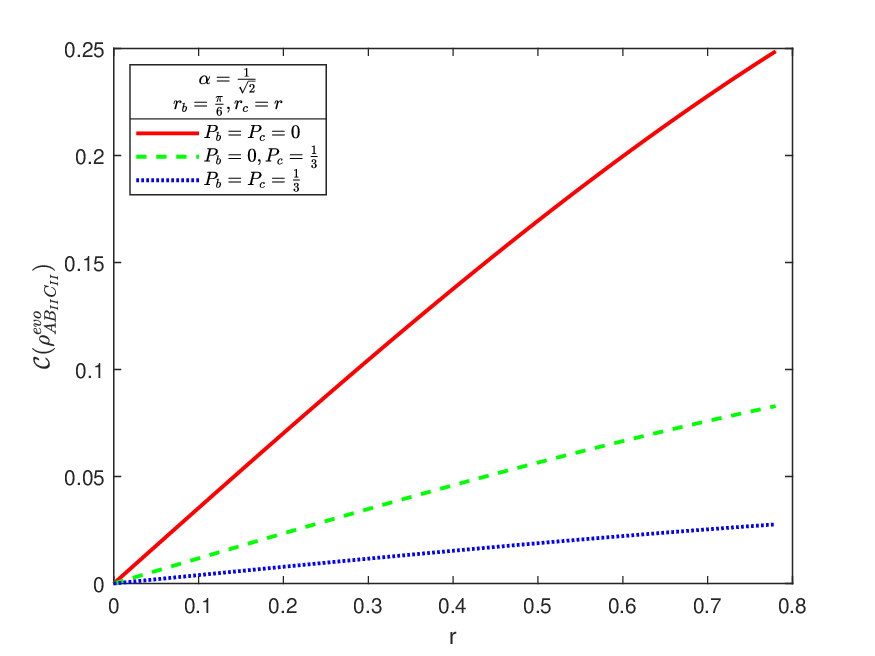}
		\caption{ }
		\label{fig:subfig1}
	\end{subfigure}
	\label{Fig.2}\caption{Plot coherence concurrence $\mathcal{C}(\rho_{AB_{I}C_{I}}^{evo})$  and $\mathcal{C}(\rho_{AB_{II}C_{II}}^{evo})$ as functions of  Hawking radiation $r$ under phase flip noise. }
\end{figure}

\begin{figure}
	\centering
\begin{subfigure}[h]{0.32\textwidth}
		\includegraphics[width=\textwidth]{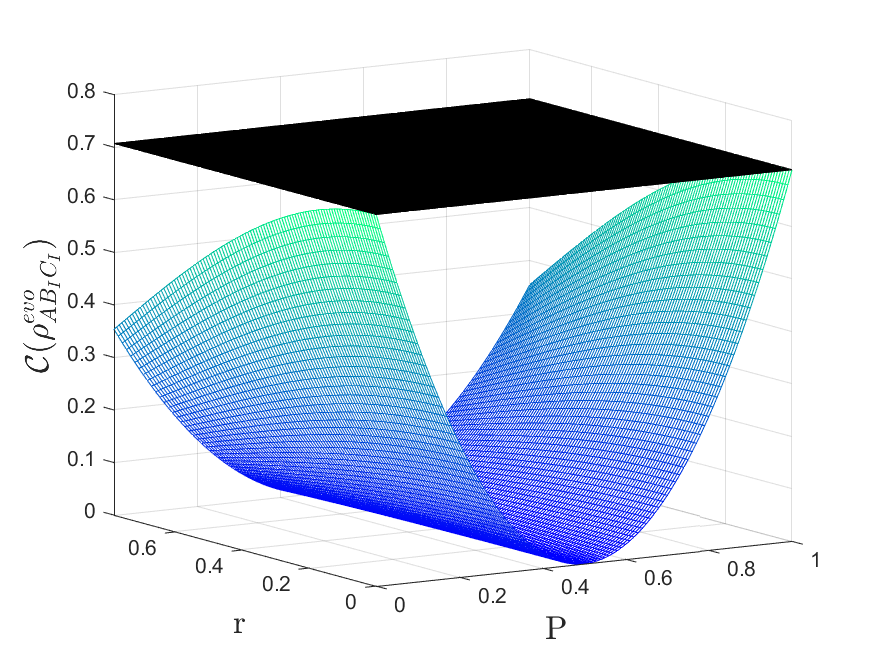}
		\caption{ }
		\label{fig:subfig1}
	\end{subfigure}
	\begin{subfigure}[h]{0.32\textwidth}
		\includegraphics[width=\textwidth]{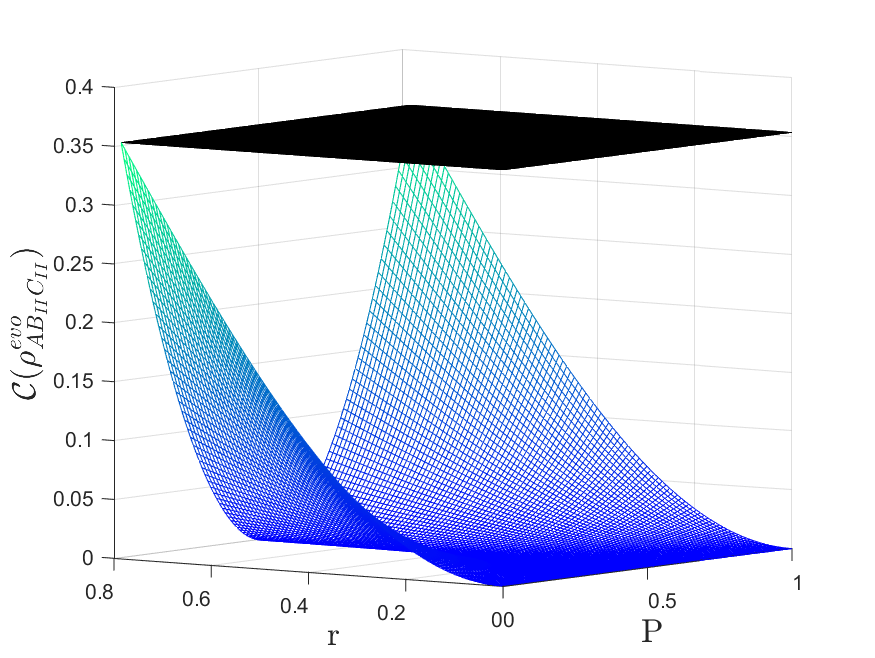}
		\caption{}
		\label{fig:subfig1}
	\end{subfigure}
  \begin{subfigure}[h]{0.32\textwidth}
		\includegraphics[width=\textwidth]{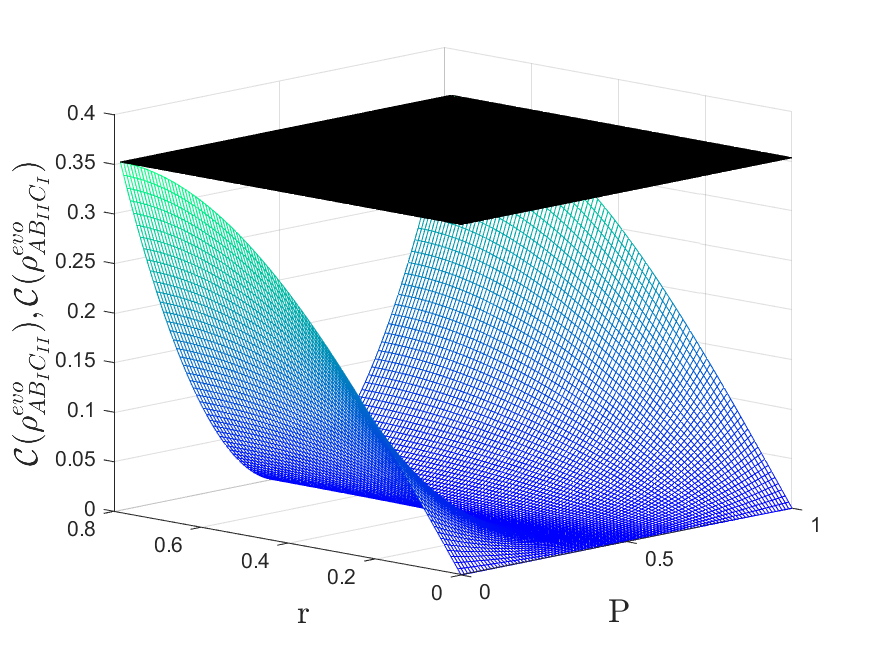}
		\caption{ }
		\label{fig:subfig1}
	\end{subfigure}
	\label{Fig.2}\caption{Plot $\mathcal{C}(\rho_{AB_{I}C_{I}}^{evo})$, $\mathcal{C}(\rho_{AB_{II}C_{II}}^{evo})$, $\mathcal{C}(\rho_{AB_{I}C_{II}}^{evo})$ under phase flip noise  when $r_b=r_c=r$, $P_b=P_c=P$ and $\alpha=\frac{1}{\sqrt{2}}$.}
\end{figure}

In Fig.4, we consider the influence of both Hawking radiation and noisy environment interactive impact simultaneously for the state parameter $\alpha=\frac{1}{\sqrt{2}}$. We find that the quantum coherence concurrence in physically accessible mode is no more than 0.7, while that in physically inaccessible mode are no more than 0.35. Meanwhile, all the coherence concurrences are decreasing at first and then increasing as the decay probability rising. Specifically, both the physically accessible and inaccessible coherence concurrence vanish completely when $P=\frac{1}{2}$.

\textbf{In case of bit flip channel}:
When we consider the case of the bit flip channel, the single qubit Kraus operators are given by,
\begin{equation}
\begin{aligned}
&E_0=\left(\begin{array}{cc}
		\sqrt{1-P} & 0 \\
		0 & \sqrt{1-P}
	\end{array}\right),
&E_1=\left(\begin{array}{cc}
		0 & \sqrt{P} \\
		\sqrt{P} & 0
	\end{array}\right).\\
\end{aligned}
\end{equation}

Under the bit flip channel, the final density matrix of $\rho_{AB_IC_I}$ evolves to
\begin{equation}
\rho^{evo}_{AB_IC_I}=\left(
  \begin{array}{cccccccc}
    a_1 & 0 & 0 & 0 & 0 & 0 & 0 & f_1 \\
    0 & a_2 & 0 & 0 & 0 & 0 & 0 & 0 \\
    0 & 0 & a_3 & 0 & 0 & 0 & 0 & 0 \\
    0 & 0 & 0 & a_4 & 0 & 0 & 0 & 0 \\
    0 & 0 & 0 & 0 & 0 & 0 & 0 & 0 \\
    0 & 0 & 0 & 0 & 0 & 0 & 0 & 0 \\
    0 & 0 & 0 & 0 & 0 & 0 & 0 & 0 \\
   f_1 & 0 & 0 & 0 & 0 & 0 & 0 & \frac{\alpha}{2} \\
  \end{array}
\right)
\end{equation}
where matrix element
\begin{equation}
f_1=\frac{\alpha(1 -P_b ) (1-P_c )}{2} \cos r_b \cos r_c.
\end{equation}

As in the previous sections, in this case we get coherence concurrence of $\rho^{evo}_{AB_IC_I}$
\begin{equation}
\mathcal{C}(\rho^{evo}_{AB_IC_I})=\alpha(1 -P_b ) (1-P_c ) \cos r_b \cos r_c.
\end{equation}
In a similarly way, we obtain the physically inaccessible coherence concurrences after influence by the bit flip channel,
\begin{equation}
\begin{aligned}
\mathcal{C}(\rho^{evo}_{AB_IC_{II}})&=\alpha(1 -P_b ) (1-P_c ) \cos r_b \sin r_c,\\
\mathcal{C}(\rho^{evo}_{AB_{II}C_{I}})&=\alpha(1 -P_b ) (1-P_c ) \sin r_b \cos r_c,\\
\mathcal{C}(\rho^{evo}_{AB_{II}C_{II}})&=\alpha(1 -P_b ) (1-P_c ) \sin r_b \sin r_c.\\
\end{aligned}
\end{equation}

\begin{figure}
	\centering
	\begin{subfigure}[h]{0.32\textwidth}
		\includegraphics[width=\textwidth]{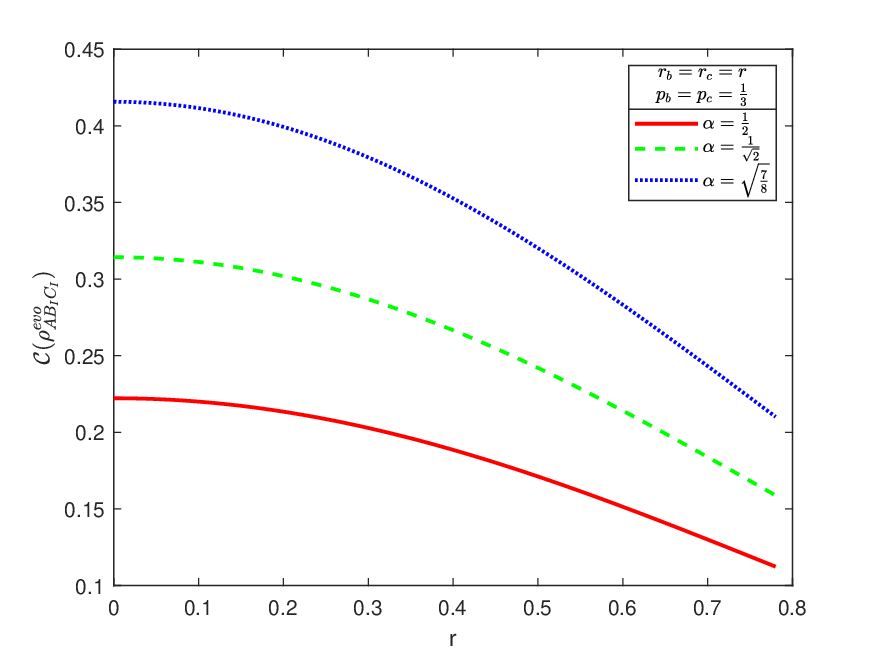}
		\caption{ }
		\label{fig:subfig1}
	\end{subfigure}
  \begin{subfigure}[h]{0.32\textwidth}
		\includegraphics[width=\textwidth]{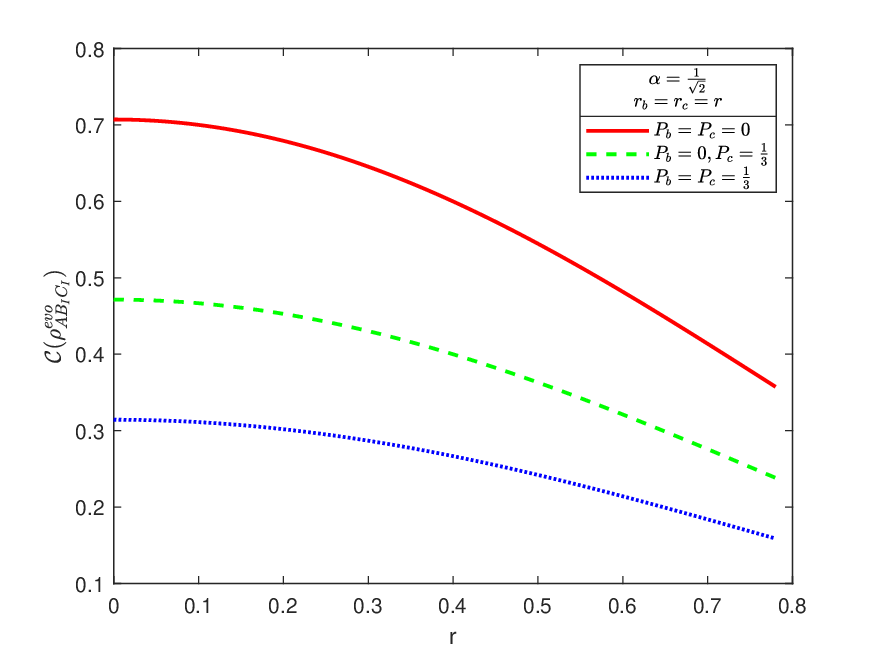}
		\caption{}
		\label{fig:subfig1}
	\end{subfigure}
 \begin{subfigure}[h]{0.32\textwidth}
		\includegraphics[width=\textwidth]{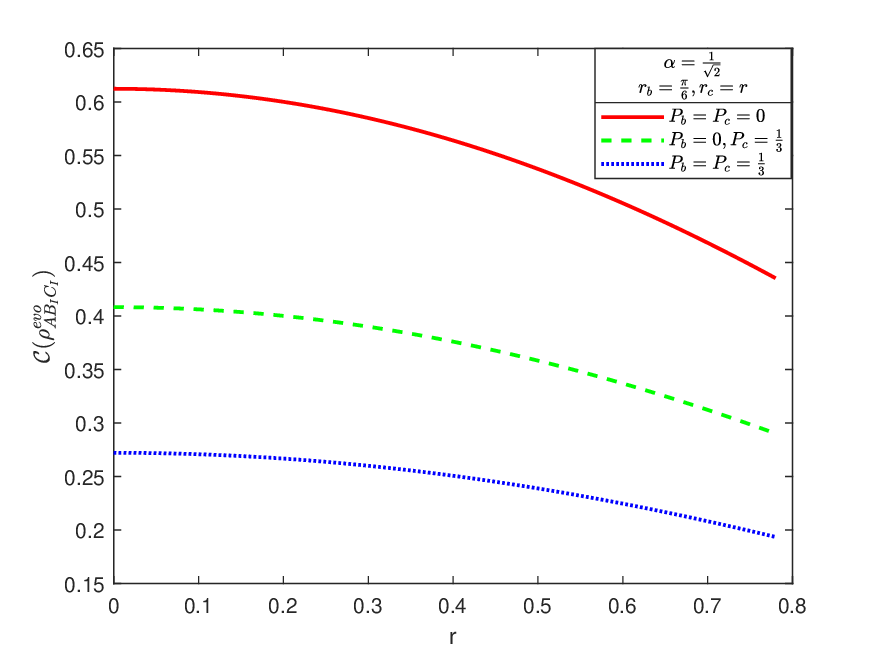}
		\caption{  }
		\label{fig:subfig1}
	\end{subfigure}
	\begin{subfigure}[h]{0.32\textwidth}
		\includegraphics[width=\textwidth]{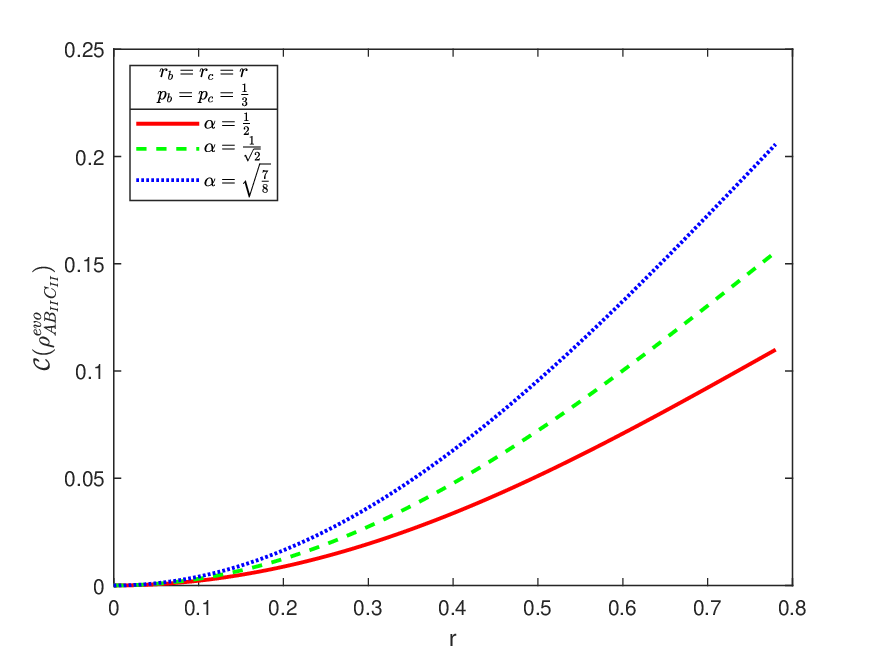}
		\caption{ }
		\label{fig:subfig1}
	\end{subfigure}
  \begin{subfigure}[h]{0.32\textwidth}
		\includegraphics[width=\textwidth]{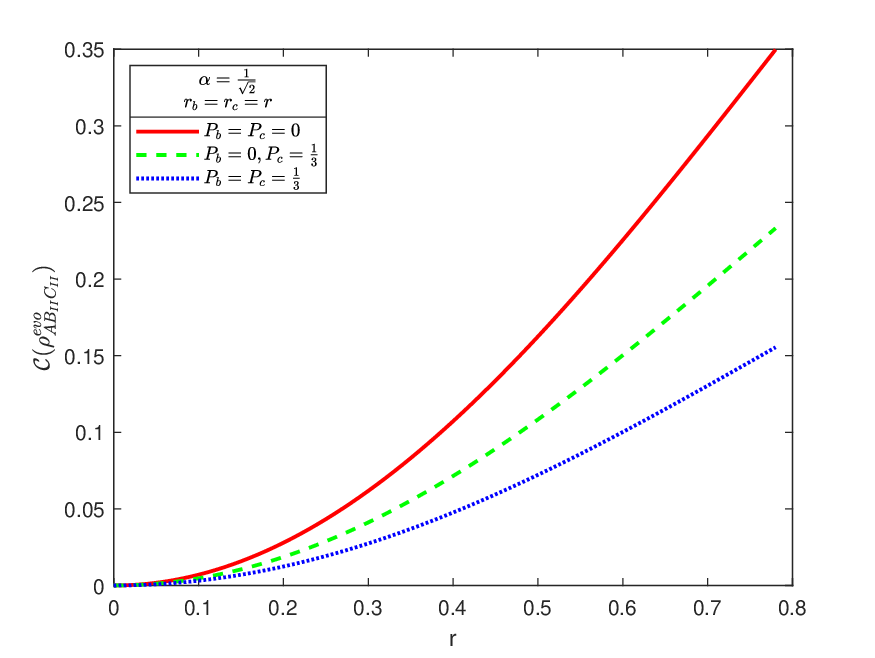}
		\caption{}
		\label{fig:subfig1}
	\end{subfigure}
 \begin{subfigure}[h]{0.32\textwidth}
		\includegraphics[width=\textwidth]{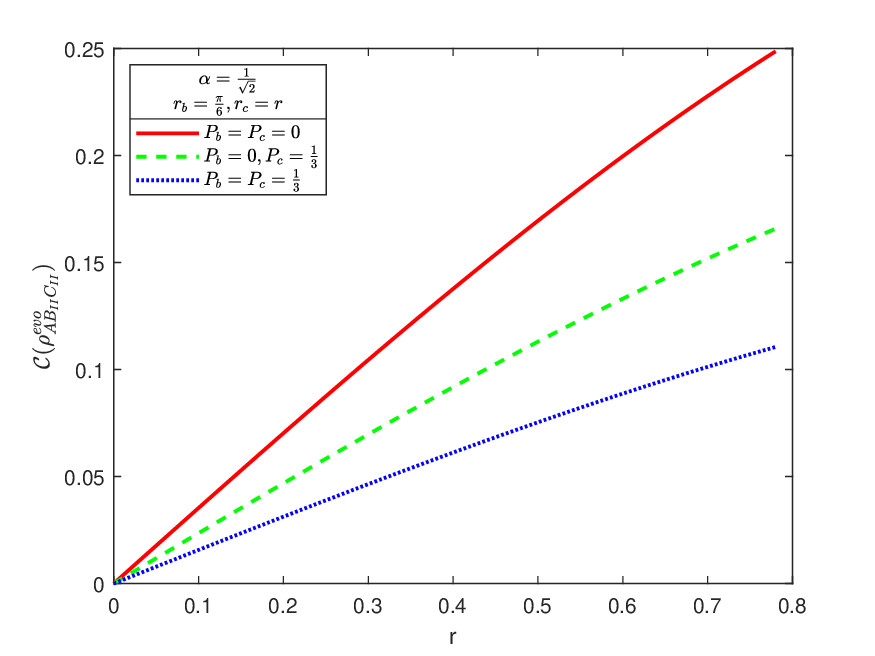}
		\caption{ }
		\label{fig:subfig1}
	\end{subfigure}
	\label{Fig.2}\caption{Plot $\mathcal{C}(\rho_{AB_{I}C_{I}}^{evo})$ and $\mathcal{C}(\rho_{AB_{II}C_{II}}^{evo})$  as functions of  Hawking radiation $r$ under bit flip noise.}
\end{figure}

\begin{figure}
	\centering
\begin{subfigure}[h]{0.32\textwidth}
		\includegraphics[width=\textwidth]{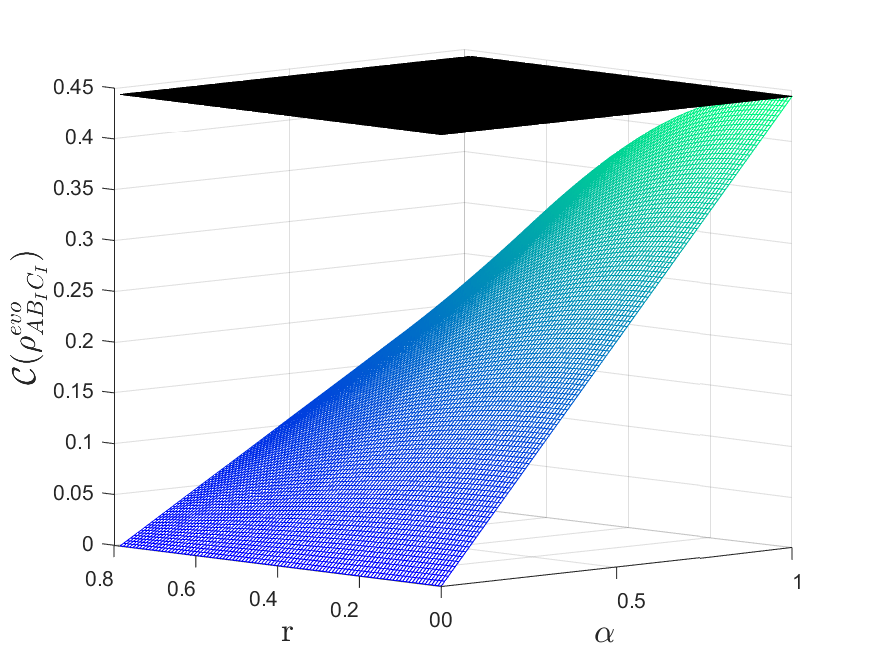}
		\caption{ }
		\label{fig:subfig1}
	\end{subfigure}
	\begin{subfigure}[h]{0.32\textwidth}
		\includegraphics[width=\textwidth]{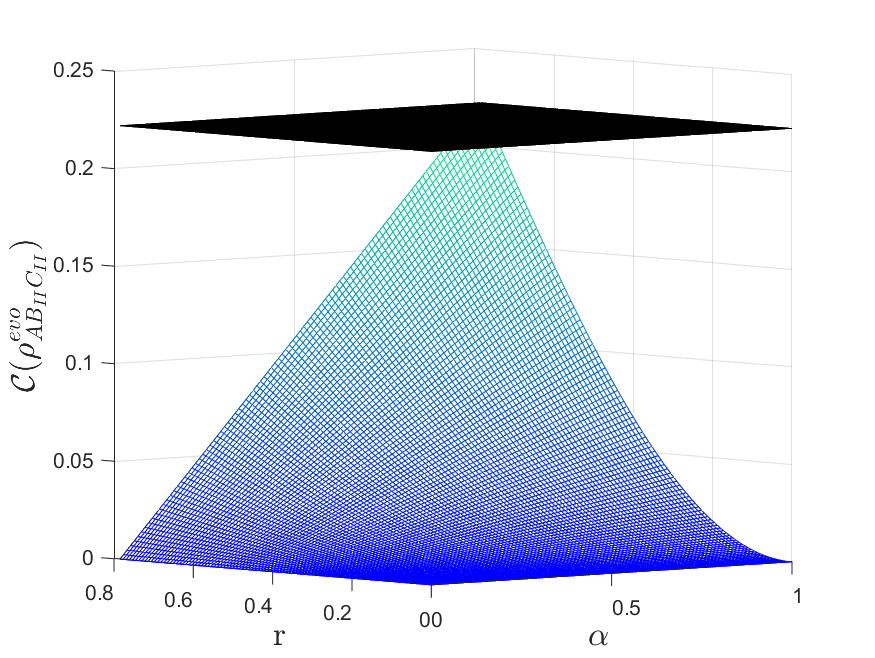}
		\caption{}
		\label{fig:subfig1}
	\end{subfigure}
  \begin{subfigure}[h]{0.32\textwidth}
		\includegraphics[width=\textwidth]{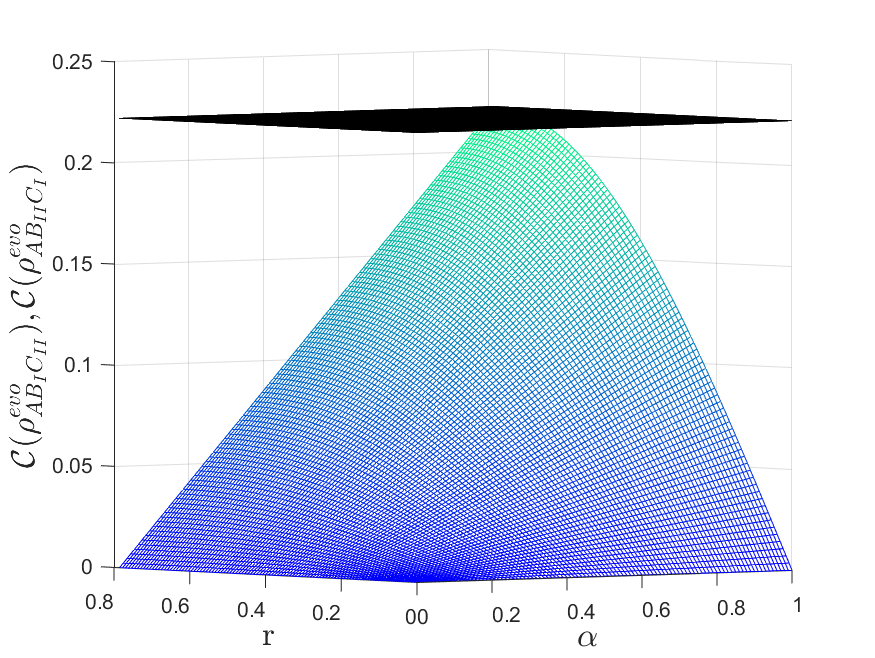}
		\caption{ }
		\label{fig:subfig1}
	\end{subfigure}
	\label{Fig.2}\caption{Plot $\mathcal{C}(\rho_{AB_{I}C_{I}}^{evo})$, $\mathcal{C}(\rho_{AB_{II}C_{II}}^{evo})$, $\mathcal{C}(\rho_{AB_{II}C_{I}}^{evo})$ under bit flip noise  when $r_b=r_c=r$, $P_b=P_c=\frac{1}{3}$.}
\end{figure}

In Fig. 5 and Fig. 6, we plot coherence concurrence as a function of acceleration with several fixed values of the $\alpha$ or decay probability $P$ for the bit flip channel. Both the noisy environment
and the acceleration of subsystems influence the coherence concurrences,
but the coherence concurrences sudden death never occurs.
The coherence concurrences are increasing as the state parameter $\alpha$ rising,  the value of physically accessible coherence concurrence is gradually approaching to 0.45 when $P_b=P_c=\frac{1}{3}$, while physically inaccessible coherence concurrence is approximately increasing to 0.225 as the state parameter increases.

\section{Conclusions}
In this paper, based on generalized GHZ mixed states with X shape as initial states, shared by Alice, Bob, and Charlie, we have studied the effect of acceleration and environment noises on tripartite
coherence of a three-qubit Dirac system when
two subsystems are accelerated.
The coherence concurrence is employed as a measure of coherence of mixed states, and which is computed numerically under three different types of noisy channels. Both the acceleration and noisy channels can modulate the behaviour of the tripartitle coherence concurrence.  The coherence concurrence can be decreased by Unruh effect, but they do not vanish even for infinite acceleration, while they can happen to reduce to zero by noisy interaction. In other words, noisy environments affect the genuine tripartite coherenc much more strongly than the acceleration of
subsystems.

Both coherence concurrence and $l_1$-norm of coherence are the measures of coherence, they are equal to each other for pure states. Generally speaking, for the qudit mixed states, $l_1$-norm of coherence is no more than the coherence concurrence. Surprisingly, in case of generalized GHZ mixed states with X shape,
we find that the two measures are tantamount again. Therefore, we may predict that coherence concurrences of all qudit states with X shape
always remain the same with the $l_1$ coherence in accelerated
frames.

\bigskip
\noindent{\bf Acknowledgments}

The research is supported by the National Natural Science Foundation of China under
Grant Nos. 12171044; the Hainan Provincial Natural Science Foundation of China under
Grant No. 121RC539; the specific research fund of the Innovation Platform for Academicians of Hainan
Province under Grant No. YSPTZX202215.

\section{Appendix A}

We calculate the  physically inaccessible quantum tripartite subsystems after acceleration as follows,
\begin{equation}\label{ab2c1}
\rho_{AB_{II}C_I}=\left(
  \begin{array}{cccccccc}
    a_1 & 0 & 0 & 0 & 0 & 0 & 0 & 0 \\
    0 & a_2 & 0 & 0 & 0 & 0 & 0 & 0 \\
    0 & 0 & a_3 & 0 & 0 & b_2 & 0 & 0 \\
    0 & 0 & 0 & a_4 & 0 & 0 & 0 & 0 \\
    0 & 0 & 0 & 0 & 0 & 0 & 0 & 0 \\
    0 & 0 & b_2 & 0 & 0 & \frac{\alpha}{2} & 0 & 0 \\
    0 & 0 & 0 & 0 & 0 & 0 & 0 & 0 \\
    0 & 0 & 0 & 0 & 0 & 0 & 0 & 0 \\
  \end{array}
\right),
\end{equation}
\begin{equation}\label{ab1c2}
\rho_{AB_{I}C_{II}}=\left(
  \begin{array}{cccccccc}
    a_1 & 0 & 0 & 0 & 0 & 0 & 0 & 0 \\
    0 & a_2 & 0 & 0 & 0 & 0 & b_3 & 0 \\
    0 & 0 & a_3 & 0 & 0 & 0 & 0 & 0 \\
    0 & 0 & 0 & a_4 & 0 & 0 & 0 & 0 \\
    0 & 0 & 0 & 0 & 0 & 0 & 0 & 0 \\
    0 & 0 & 0 & 0 & 0 & 0& 0 & 0 \\
    0 & b_3 & 0 & 0 & 0 & 0 & \frac{\alpha}{2}  & 0 \\
    0 & 0 & 0 & 0 & 0 & 0 & 0 & 0 \\
  \end{array}
\right),
\end{equation}
\begin{equation}\label{ab2c2}
\rho_{AB_{II}C_{II}}=\left(
  \begin{array}{cccccccc}
    a_1 & 0 & 0 & 0 & 0 & 0 & 0 & 0 \\
    0 & a_2 & 0 & 0 & 0 & 0 & 0 & 0 \\
    0 & 0 & a_3 & 0 & 0 & 0 & 0 & 0 \\
    0 & 0 & 0 & a_4 & b_4 & 0 & 0 & 0 \\
    0 & 0 & 0 & b_4 & \frac{\alpha}{2} & 0 & 0 & 0 \\
    0 & 0 & 0 & 0 & 0 & 0& 0 & 0 \\
    0 & 0 & 0 & 0 & 0 & 0 & 0  & 0 \\
    0 & 0 & 0 & 0 & 0 & 0 & 0 & 0 \\
  \end{array}
\right),
\end{equation}
\begin{equation}\label{ab1b2}
\rho_{AB_IB_{II}}=\left(
  \begin{array}{cccccccc}
    c_1 & 0 & 0 & 0 & 0 & 0 & 0 & b_5 \\
    0 & c_2 & 0 & 0 & 0 & 0 & 0 & 0 \\
    0 & 0 & c_2 & 0 & 0 & 0 & 0 & 0 \\
    0 & 0 & 0 & c_3 & 0 & 0 & 0 & 0 \\
    0 & 0 & 0 & 0 & 0 & 0 & 0 & 0 \\
    0 & 0 & 0 & 0 & 0 & 0 & 0 & 0 \\
    0 & 0 & 0 & 0 & 0 & 0 & 0 & 0 \\
    b_5 & 0 & 0 & 0 & 0 & 0 & 0 & \frac{\alpha}{2} \\
  \end{array}
\right)
\end{equation}
\begin{equation}\label{ab1c1}
\rho_{AC_IC_{II}}=\left(
  \begin{array}{cccccccc}
    g_1 & 0 & 0 & 0 & 0 & 0 & 0 & b_6 \\
    0 & g_2 & 0 & 0 & 0 & 0 & 0 & 0 \\
    0 & 0 & g_2 & 0 & 0 & 0 & 0 & 0 \\
    0 & 0 & 0 & g_3 & 0 & 0 & 0 & 0 \\
    0 & 0 & 0 & 0 & 0 & 0 & 0 & 0 \\
    0 & 0 & 0 & 0 & 0 & 0 & 0 & 0 \\
    0 & 0 & 0 & 0 & 0 & 0 & 0 & 0 \\
    b_6 & 0 & 0 & 0 & 0 & 0 & 0 & \frac{\alpha}{2} \\
  \end{array}
\right)
\end{equation}
where the matrix elements
\begin{equation}
\begin{aligned}
b_2&=\frac{\alpha}{2}\sin r_b \cos r_c,\\
b_3&=\frac{\alpha}{2}\cos r_b \sin r_c,\\
b_4&=\frac{\alpha}{2}\sin r_b \sin r_c,\\
b_5&=\frac{\alpha}{2}\cos^2 r_b ,\\
b_6&=\frac{\alpha}{2}\cos^2 r_c .\\
\end{aligned}
\end{equation}
\begin{equation}
\begin{aligned}
c_1&=\frac{2-\alpha}{2}\cos^4 r_b,\\
c_2&=\frac{2-\alpha}{2}\cos^2 r_b \sin^2 r_b,\\
c_3&=\frac{2-\alpha}{2}\sin^4 r_b ,\\
\end{aligned}
\end{equation}
\begin{equation}
\begin{aligned}
g_1&=\frac{2-\alpha}{2}\cos^4 r_c,\\
g_2&=\frac{2-\alpha}{2}\cos^2 r_c \sin^2 r_c,\\
g_3&=\frac{2-\alpha}{2}\sin^4 r_c ,\\
\end{aligned}
\end{equation}

\section{Appendix B}
Under effect of acceleration and the phase damping noisy, the tripartite reduce states $\rho_{AB_IC_I}$, $\rho_{AB_{I}C_{II}}$, $\rho_{AB_{II}C_I}$ and $\rho_{AB_{II}C_{II}}$ evolve to those states as follows, respectively,
\begin{equation}
\rho^{evo}_{AB_IC_I}=\left(
  \begin{array}{cccccccc}
    a_1 & 0 & 0 & 0 & 0 & 0 & 0 & d_1 \\
    0 & a_2 & 0 & 0 & 0 & 0 & 0 & 0 \\
    0 & 0 & a_3 & 0 & 0 & 0 & 0 & 0 \\
    0 & 0 & 0 & a_4 & 0 & 0 & 0 & 0 \\
    0 & 0 & 0 & 0 & 0 & 0 & 0 & 0 \\
    0 & 0 & 0 & 0 & 0 & 0 & 0 & 0 \\
    0 & 0 & 0 & 0 & 0 & 0 & 0 & 0 \\
    d_1 & 0 & 0 & 0 & 0 & 0 & 0 & \frac{\alpha}{2} \\
  \end{array}
\right),
\end{equation}
\begin{equation}
\rho^{evo}_{AB_{II}C_I}=\left(
  \begin{array}{cccccccc}
    a_1 & 0 & 0 & 0 & 0 & 0 & 0 & 0 \\
    0 & a_2 &0 & 0 & 0 & 0 & 0 & 0 \\
    0 & 0 & a_3 & 0 & 0 & 0 & d_2 & 0 \\
    0 & 0 & 0 & a_4 &0 & 0 & 0 & 0 \\
    0 & 0 & 0 & 0 & 0 & 0 & 0 & 0 \\
    0 & 0 & 0 & 0 & 0 & 0 & 0 & 0 \\
    0 & 0 & d_2 & 0 & 0 & 0 & \frac{\alpha}{2} & 0 \\
    0 & 0 & 0 & 0 & 0 & 0 & 0 & 0 \\
  \end{array}
\right),
\end{equation}
\begin{equation}
\rho^{evo}_{AB_{I}C_{II}}=\left(
  \begin{array}{cccccccc}
    a_1 & 0 & 0 & 0 & 0 & 0 & 0 & 0 \\
    0 & a_2 & 0 & 0 & 0 & 0 & d_3 & 0 \\
    0 & 0 & a_3 & 0 & 0 & 0 & 0 & 0 \\
    0 & 0 & 0 & a_4 & 0 & 0 & 0 & 0 \\
    0 & 0 & 0 & 0 & 0 & 0 & 0 & 0 \\
    0 & 0 & 0 & 0 & 0 & 0& 0 & 0 \\
    0 & d_3 & 0 & 0 & 0 & 0 & \frac{\alpha}{2}  & 0 \\
    0 & 0 & 0 & 0 & 0 & 0 & 0 & 0 \\
  \end{array}
\right),
\end{equation}
\begin{equation}
\rho^{evo}_{AB_{II}C_{II}}=\left(
  \begin{array}{cccccccc}
    a_1 & 0 & 0 & 0 & 0 & 0 & 0 & 0 \\
    0 & a_2 & 0 & 0 & 0 & 0 & 0 & 0 \\
    0 & 0 & a_3 & 0 & 0 & 0 & 0 & 0 \\
    0 & 0 & 0 & a_4 & d_4 & 0 & 0 & 0 \\
    0 & 0 & 0 & d_4 & \frac{\alpha}{2} & 0 & 0 & 0 \\
    0 & 0 & 0 & 0 & 0 & 0& 0 & 0 \\
    0 & 0 & 0 & 0 & 0 & 0 & 0  & 0 \\
    0 & 0 & 0 & 0 & 0 & 0 & 0 & 0 \\
  \end{array}
\right),
\end{equation}
where
\begin{equation}
\begin{aligned}
d_1&=\frac{\alpha}{2}\sqrt{1-p_b}\sqrt{1-p_c}\cos r_b \cos r_c,\\
d_2&=\frac{\alpha}{2}\sqrt{1-p_b}\sqrt{1-p_c}\sin r_b \cos r_c,\\
d_3&=\frac{\alpha}{2}\sqrt{1-p_b}\sqrt{1-p_c}\cos r_b \sin r_c,\\
d_4&=\frac{\alpha}{2}\sqrt{1-p_b}\sqrt{1-p_c}\sin r_b \sin r_c.
\end{aligned}
\end{equation}

\end{document}